\documentclass[a4paper,11pt]{article}
\usepackage{pos}
\usepackage{slashed}
\usepackage{dsfont}
\usepackage{euscript}
\usepackage{xcolor}
\usepackage{hyperref}
\usepackage{doi}
\definecolor{mypink}{RGB}{255,105,180}

\title{Twist-3 T-Even TMD Distributions in Quark Target Model}

\author*[a]{Siddhesh Padval}
\author[b]{Anuradha Misra}
\author[c]{Abhirup Karmakar}
\author[d]{Deepesh Bhamre}

\affiliation[a]{Department of Physics, 
University of Mumbai, Santacruz (East), Mumbai-400098, India}

\affiliation[b]{Centre for Excellence in Basic Sciences (UMDAE-CEBS), 
University of Mumbai, Santacruz (East), Mumbai-400098, India}

\affiliation[c]{School of Physical Sciences, Indian Association for the Cultivation of Science, 2A and 2B Raja S.C.
Mullick Road, Kolkata 700 032, India}

\affiliation[d]{Laborat\'{o}rio de F\'{i}sica Te\'{o}rica e Computacional-LFTC, Universidade Cruzeiro do Sul/Universidade Cidade de S\~{a}o Paulo (UNICID), Rua Galv\~{a}o Bueno, 01506-000, S\~{a}o Paulo, Brazil}

\emailAdd{siddhesh.padval@physics.mu.ac.in}

\abstract{
Transverse momentum dependent parton distributions are objects that have gained substantial attention in the recent past as they are crucial for studying the 3-D structure of proton.
These distributions at twist-3 are important for understanding phenomena like beam single spin asymmetries, which are observed in high-energy scattering experiments.
However, due to the unavailability of direct phenomenological fits, currently one has to rely on simplified models of the proton in order to obtain these twist-3 distributions. 
In this work, we present the calculation of T-even twist-3 quark distributions using the quark target model.
}

\FullConference{17th International Symposium on Radiative Corrections: Applications of Quantum Field Theory to Phenomenology (RADCOR2025)\\
5-10 October 2025\\
Puri, India\\}

\begin{document}

\maketitle

\section{Introduction}
Twist-3 distributions have attracted significant attention following recent measurements of Beam Single Spin Asymmetries (BSSAs) in semi-inclusive deep-inelastic scattering (SIDIS) processes.
A recent example of these measurements is the BSSAs reported by the CLAS collaboration in charged kaon electroproduction off protons in the valence region \cite{CLAS:2025asy}.
The twist-3 distributions have not yet been directly extracted or fitted from experimental data, unlike twist-2 distributions such as unpolarized collinear parton distribution functions (PDFs), transverse momentum dependent (TMD) PDFs and the Sivers distribution, which are available. 
Therefore, in order to provide theoretical predictions for BSSAs using twist-3 distributions, one must rely on simplified models of the proton and compute these distributions within such frameworks.
So far, the quark-diquark spectator model has been widely used in the literature for estimating BSSAs \cite{Mao:2012dk, Mao:2013waa}. 
However, due to the absence of explicit gluon degrees of freedom in the spectator quark-diquark model, gluon distributions cannot be computed within this framework.
In this work, we aim to provide an alternative approach by employing the Quark Target Model (QTM) \cite{Costa:2021mpk, Meissner:2007rx} to calculate twist-3 distributions. 

In the QTM, a hadron is treated as consisting of a single dressed quark, and the corresponding Lagrangian reduces to the QCD Lagrangian for the special case of one quark flavor \cite{Meissner:2007rx}.
This framework naturally incorporates gluonic degrees of freedom.
Although the distributions calculated in QTM cannot be directly used for SIDIS phenomenology, it has been shown in the literature that appropriate convolutions of QTM distributions with quark-in-hadron distributions allow one to obtain realistic hadronic distributions \cite{Costa:2021mpk}. 
This procedure requires an additional low-energy QCD input, for instance, from models such as the NJL model consisting of only quark degrees of freedom.
In this work, we present our calculations of the twist-3 T-even quark distributions within QTM.

In Sec.\ref{sec:formalism}, we briefly discuss the formalism within which the distributions are calculated.
The expressions for twist-3 T-even quark distributions are provided in Sec.\ref{sec:distributions}.
Finally, in Sec.\ref{sec:conclusion}, we conclude with a few remarks about the application of the results obtained in this work for a future study. 

\section{Formalism}\label{sec:formalism}

In the quark target model, the target state is composed of a single dressed quark, and the required parton (quark/gluon) distributions are calculated perturbatively in QCD.
The Lagrangian of QTM is given by \cite{Meissner:2007rx}

\begin{equation}\label{eq:QTM_lagrangian}
\mathcal L_{QTM} = \bar\psi(x)(i\gamma^\mu D_\mu - m)\psi(x) - \frac{1}{4}F^{\mu\nu}_a(x)F_{\mu\nu,a}(x)
\end{equation}
where the gluon field strength tensor and the covariant derivative are defined as
\begin{align}
F^{\mu\nu}_a(x) &= \partial^\mu A^\nu_a(x) - \partial^\nu A^\mu_a(x) + gf_{abc}A^\mu_b(x)A^\nu_c(x)\\
D^\mu\psi(x) &= [\partial^\mu - igt_aA^\mu_a(x)]\psi(x)
\end{align}

Note that although $\mathcal L_{QTM}$ of eq.(\ref{eq:QTM_lagrangian}) looks the same as the QCD Lagrangian, the fields appearing in $\mathcal L_{QTM}$ are dressed by the interactions of some low-energy effective theory of QCD.
Hence, the mass $m$ appearing in eq.(\ref{eq:QTM_lagrangian}) too is not the current quark mass, but a dressed quark mass.
As can be seen, the three-gluon vertex enters naturally into the computation of distribution functions in this model.
This is a useful feature of QTM which is unavailable in some other simpler models, like the scalar diquark model for instance.

We consider a dressed quark of mass $M$, spin $S$ and 4-momentum $P$ composed of a bare quark of momentum $k$ and a gluon of momentum $(P-k)$.
We set the momentum of the bare quark as $k = (xP^+,k^-,\mathbf k_\perp)$, where the longitudinal momentum fraction $x$ is defined as $x = k^+/P^+$.
We can choose a coordinate system where the incoming dressed quark has zero transverse momentum, i.e., $P = (P^+,P^-,\mathbf 0_\perp)$, where $P^- = M^2/2P^+$ due to the on-shell condition.
Respecting the standard relations $P\cdot S =0,\; S^2 = -1$, we set our convention for the spin 4-vector \cite{Bacchetta:2006tn} as $S = (S_zP^+/M, -S_zM/2P^+, \mathbf S_\perp)$.
We also introduce the lightlike vectors $n_+ = (1,0,\mathbf 0_\perp)$ and $n_- = (0,1,\mathbf 0_\perp)$, the antisymmetric tensor $\epsilon_\perp^{12} = - \epsilon_\perp^{21} = 1$ with all other components zero, and the transverse metric tensor $g_\perp^{11} = g_\perp^{22} = -1$ with all other components zero.

The quark-quark correlator, upto twist-3, is given by \cite{Bacchetta:2006tn}

\begin{equation}\label{eq:correlator}
\begin{aligned}
\Phi(x,\mathbf k_\perp)
=& 
\frac{1}{2} \Bigg\{
f_{1} \slashed{n}_{+} - f_{1T}^\perp\frac{\epsilon_T^{\rho\sigma} k_{\perp\rho} S_{\perp\sigma}}{M} \slashed{n}_{+} + g_{1s} \gamma_{5} \slashed{n}_{+}
\\
&\phantom{\frac{1}{2} \Bigg\{ }+ h_{1T} \frac{[\slashed{S}_{\perp},\slashed{n}_{+}] \gamma_{5}}{2} + h_{1s}^{\perp} \frac{[\slashed{k}_{\perp},\slashed{n}_{+}] \gamma_{5}}{2M} + ih_{1}^{\perp} \frac{[\slashed{k}_{\perp},\slashed{n}_{+}]}{2M} \Bigg\}
\\
& 
+ \frac{M}{2P^+} \Bigg\{ 
e - i e_{s} \gamma_{5} - e_{T}^{\perp}\frac{\epsilon_{T}^{\rho\sigma} k_{\perp\rho} S_{\perp\sigma}}{M}
\\
&\phantom{+ \frac{M}{2P^+} \Bigg\{ } + f^{\perp} \frac{\slashed{k}_{\perp}}{M} - f'_{T} \epsilon_{T}^{\rho\sigma} \gamma_{\rho} S_{\perp\sigma} - f_{s}^{\perp} \frac{\epsilon_{T}^{\rho\sigma} \gamma_{\rho} k_{\perp\sigma}}{M}
\\
&\phantom{+ \frac{M}{2P^+} \Bigg\{ } + g'_{T} \gamma_{5} \slashed{S}_{\perp} + g_{s}^{\perp} \gamma_{5} \frac{\slashed{k}_{\perp}}{M} - g^{\perp} \gamma_5\frac{\epsilon_{T}^{\rho\sigma} \gamma_{\rho} k_{\perp\sigma}}{M}
\\
&\phantom{+ \frac{M}{2P^+} \Bigg\{ } + h_{s} \frac{[\slashed{n}_{+},\slashed{n}_{-}] \gamma_{5}}{2} + h_{T}^{\perp} \frac{[\slashed{S}_{\perp},\slashed{k}_{\perp}] \gamma_{5}}{2M} + ih \frac{[\slashed{n}_{+},\slashed{n}_{-}]}{2} \Bigg\}
\end{aligned}
\end{equation}
Here, we have used the shorthand notation \cite{Bacchetta:2006tn}
\begin{equation}
    g_{1s}(x,\mathbf k_\perp^2)
    =
    S_{z}g_{1L}(x,\mathbf k_\perp^2) - \frac{\mathbf{k}_{\perp}\cdot \mathbf{S}_{\perp}}{M}g_{1T}(x,\mathbf k_\perp^2)
\end{equation}
The eight distributions appearing in the first bracket of eq.(\ref{eq:correlator}) are twist-2 distributions, whereas the terms in the second bracket contain twist-3 distributions.

\section{Twist-3 T-even quark TMD PDFs}\label{sec:distributions}

\begin{figure}
    \centering
    \includegraphics[width=0.8\linewidth]{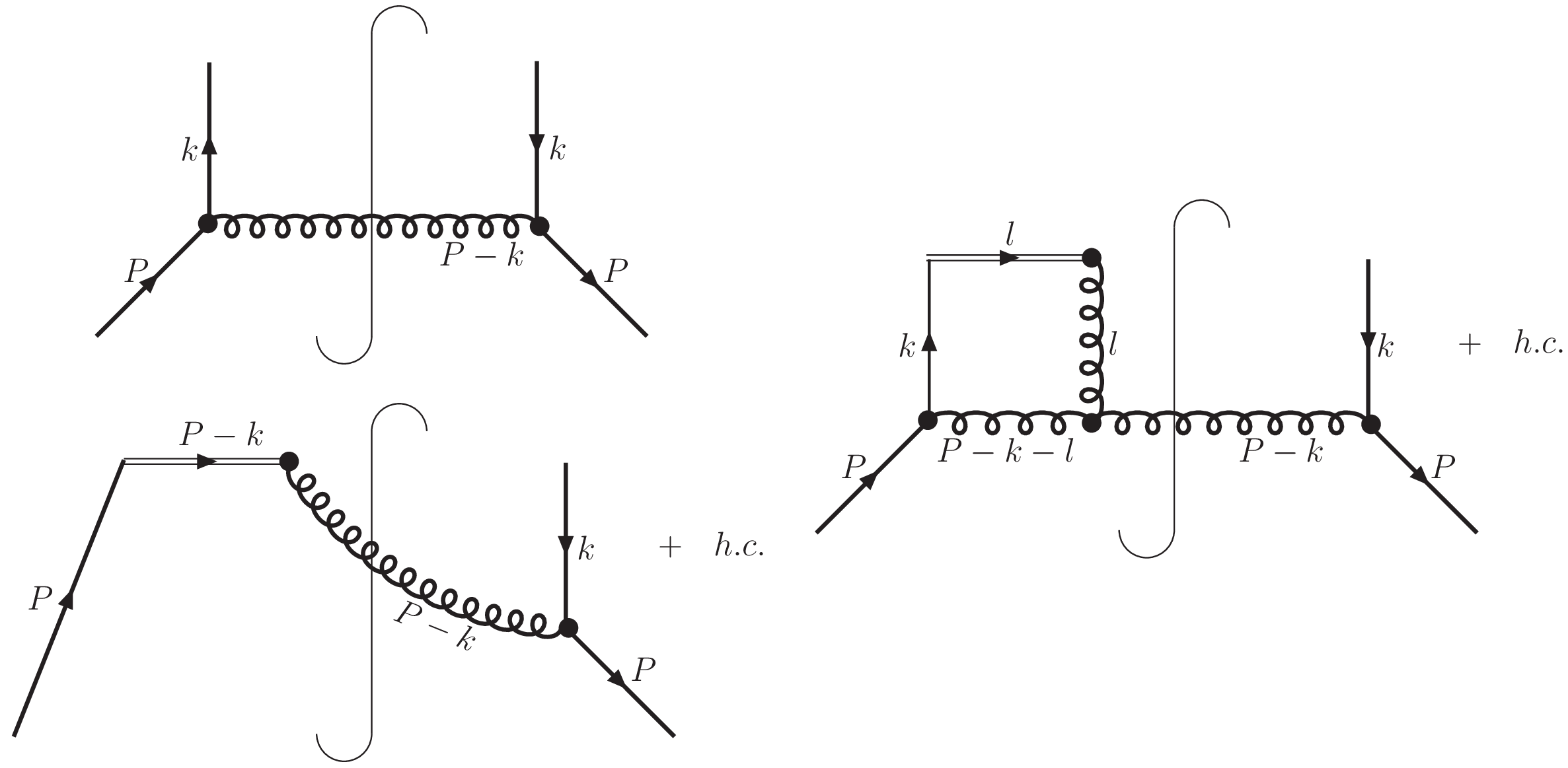}
    \caption{Handbag diagrams corresponding to the twist-3 quark-quark correlator in QTM}
    \label{fig:quark_target_handbag}
\end{figure}

    The twist-2 TMD PDFs in QTM have been calculated by Mei{\ss}ner {\it et al.} in Ref.\cite{Meissner:2007rx}.
In this work, we calculate the twist-3 T-even quark distributions in a dressed quark in QTM.
The handbag diagrams relevant to the distributions calculated here are given in Fig.(\ref{fig:quark_target_handbag}a) and (\ref{fig:quark_target_handbag}b).
At the lowest perturbative order, Figs.(\ref{fig:quark_target_handbag}a) and (\ref{fig:quark_target_handbag}b) contribute to the distributions calculated below, with Fig.(\ref{fig:quark_target_handbag}b) containing the contribution from a higher-order Wilson line.
Fig.(\ref{fig:quark_target_handbag}c) is an example of a higher perturbative order diagram, and is relevant for T-odd distributions, although not relevant in the current work.

Adopting the methodology provided in Ref.\cite{Bacchetta:2015qka} for calculating the expressions for these correlator diagrams, we get,
\\
\begin{align}
    \label{eq:correlator diag 1a}
    \Phi^{(1 a)}(k;S)_{ij} 
    &=
    \frac{1}{\left(2\pi\right)^{3}}\delta\left( \left( P - k \right)^2 \right)
    g_{s}^{2}
    \mathcal C_{F}
    \bar{u}_{k}\left(P, S\right)
    \left[
    \gamma_{\mu}
    \frac{\left( \slashed{k} + m \right)}{k^{2} -m^{2} -i\epsilon}
    \right]_{kj}
    \left[
    \frac{\left( \slashed{k} + m \right)}{k^{2} -m^{2} -i\epsilon}
    \gamma^{\mu}
    \right]_{il}
    {u}_{l}\left(P, S\right),
    \\
    \label{eq:correlator diag 1b}
    \Phi^{(1 b)}(k;S)_{ij} 
    &=
    \frac{1}{\left(2\pi\right)^{3}}\delta\left( \left( P - k \right)^2 \right)
    g_{s}^{2}
    \mathcal C_{F}
    \bar{u}_{k}\left(P, S\right)
    \left[
    \gamma^{+}
    \frac{\left( \slashed{k} + m \right)}{\left(\left(P-k\right)^{+} + i\epsilon\right)\left(k^{2} -m^{2} -i\epsilon\right)} 
    \right]_{kj}
    {u}_{i}\left(P, S\right)
\end{align}
\\
for the first two diagrams in Fig.(\ref{fig:quark_target_handbag}) respectively. 
The required distributions are now obtained by taking the appropriate Dirac $\gamma$-projections of the correlators given in eqs.(\ref{eq:correlator diag 1a}) and (\ref{eq:correlator diag 1b}).
For example, the $e(x,\mathbf k_\perp^2)$ distribution given in eq.(\ref{eq:e dist}) below can be extracted by taking the trace of $\Phi(x,\mathbf k_\perp^2)\mathds{1}$, the $f^\perp(x,\mathbf k_\perp^2)$ distribution in eq.(\ref{eq:fperp dist}) by taking the trace of $\Phi(x,\mathbf k_\perp^2)\gamma^{\alpha}$, etc.
Listed below are the twist-3 T-even quark TMD distributions that are obtained as a result of this calculation.

\begin{equation}\label{eq:e dist}
e(x,\mathbf k_\perp^2) = \frac{\mathcal C_{F}g_s^2}{(2\pi)^3}\frac{2\mathbf k_\perp^2}{(1-x)[m^2(1-x)^2+\mathbf k_\perp^2]^2}
\end{equation}

\begin{equation}\label{eq:fperp dist}
f^\perp(x,\mathbf k_\perp^2) = \frac{\mathcal C_{F}g_s^2}{(2\pi)^3}\frac{x\mathbf k_\perp^2 + m^2(1-x)^2(x-2)}{(1-x)[m^2(1-x)^2 + \mathbf k_\perp^2]^2}
\end{equation}

\begin{equation}
g_{T}(x,\mathbf k_\perp^2) = \frac{\mathcal C_{F}g_s^2}{(2\pi)^3}\frac{2\mathbf k_\perp^2}{(1-x)[m^2(1-x)^2+\mathbf k_\perp^2]^2}
\end{equation}

\begin{equation}
g_L^\perp(x,\mathbf k_\perp^2) = \frac{\mathcal C_{F}g_s^2}{(2\pi)^3}\frac{x\mathbf k_\perp^2 - m^2(1-x)^2(x-2)}{(1-x)[m^2(1-x)^2+\mathbf k_\perp^2]}
\end{equation}

\begin{equation}
h_L(x,\mathbf k_\perp^2) = \frac{\mathcal C_{F}g_s^2}{(2\pi)^3}\frac{2\mathbf k_\perp^2}{(1-x)[m^2(1-x)^2+\mathbf k_\perp^2]^2}
\end{equation}

\begin{equation}
h_T(x,\mathbf k_\perp^2) = -\frac{\mathcal C_{F}g_s^2}{(2\pi)^3}\frac{1}{(1-x)[m^2(1-x)^2+\mathbf k_\perp^2]}
\end{equation}

\begin{equation}
h_T^\perp(x,\mathbf k_\perp^2) = \frac{\mathcal C_{F}g_s^2}{(2\pi)^3}\frac{\mathbf k_\perp^2 - m^2(1-x)^2}{(1-x)[m^2(1-x)^2+\mathbf k_\perp^2]}
\end{equation}
It is to be noted that certain distributions, like $e$, $g_{T}$ and $h_{L}$, are given by the same expression.

\section{Conclusion}\label{sec:conclusion}
In this work, we have derived the expressions for twist-3 T-even TMD distributions of a quark within a dressed quark in the quark target model.
The SIDIS TMD correlator is considered at twist-3 and its various $\gamma$-projections are taken for each diagram that contributes to T-even distributions.
This leads to the expressions for the corresponding TMD distribution functions that are listed in this work.
The quark-in-dressed quark distributions having thus obtained, the actual quark-in-hadron distributions can now be calculated via a convolution of these distributions with those from a low-energy effective model, like the NJL model, of QCD.
Further, phenomenological studies of BSSAs can then be performed.
The results of this future work will be presented elsewhere.

\section*{Acknowledgements}

S. P. would like to thank the Department of Atomic Energy, Government of India for support through the Raja Ramanna Fellowship.
D. B. acknowledges the support received from Conselho Nacional de Desenvolvimento Cient\'{i}fico e Tecnol\'{o}gico (CNPq), Brasil, Process No. 152348/2024-7.
A. K. would like to thank the National Initiative on Undergraduate Science (NIUS), supported by the Department of Atomic Energy, Govt. of India, for financial support. 
A. M. would like to thank the Department of Atomic Energy, Govt. of India, for the award of Raja Ramanna Chair.

\end{document}